\documentclass{elsart3}
\usepackage{graphicx}
\usepackage{amssymb}
\begin{document}
\begin{frontmatter}

\title{Self-cooling cryogenic microcalorimeters made of SINIS junctions}
\author{Miha Furlan\corauthref{cor}},
\ead{miha.furlan@psi.ch}
\author{Eugenie Kirk, and Alex Zehnder}
\corauth[cor]{Corresponding author. Tel.: +41-56-310-4519.}

\address{Paul Scherrer Institute, Laboratory for Astrophysics,
5232 Villigen PSI, Switzerland}

\begin{abstract}
High quality low leakage SINIS devices made of Al-AlMn-Al layers were
fabricated for energy dispersive single photon detection.
Information on different heat flow channels was extracted from the
measured dynamics of detector signals due to X-ray events.
At the optimum operation point, the extracted
effective electron temperature decreased from $88 \, \mathrm{mK}$
down to $43 \, \mathrm{mK}$ due to self-cooling,
roughly doubling the detector sensitivity.
\end{abstract}

\begin{keyword}
Microcalorimeter \sep Cryogenic detector \sep
Tunnel junction \sep Microfabrication \sep Self-cooling
\PACS 85.25.Oj \sep 85.25.Am \sep 74.25.Fy \sep 74.62.Dh
\end{keyword}
\end{frontmatter}

Energy resolving single photon detectors are of great interest
for astrophysics, material science, medical applications or any
other field, where high quantum efficiency paired
with direct energy information are desirable.
Cryogenic microcalorimeters based on
normal metal--insulator--superconductor (NIS) junctions
are very attractive and
proven candidates for such high resolution detectors \cite{Nahum1995}.
Essential requirements for NIS spectrometers are low barrier
leakage currents and operation at low temperatures for a
small heat capacity of the absorber.
The latter can be met or improved by the Peltier-like effect
of hot-electron tunneling
\cite{Edwards1993,Nahum1994,Golubev2001}.
The efficiency of power transfer out of the absorber is
increased by coupling the normal metal symmetrically via
two tunnel junctions in series \cite{Leivo1996} (SINIS structure).
Heat flow  mechanisms upon energy deposition are reflected by
detector signal dynamics.

\begin{figure}[h]
   \centering
   \includegraphics[width=0.85\linewidth]{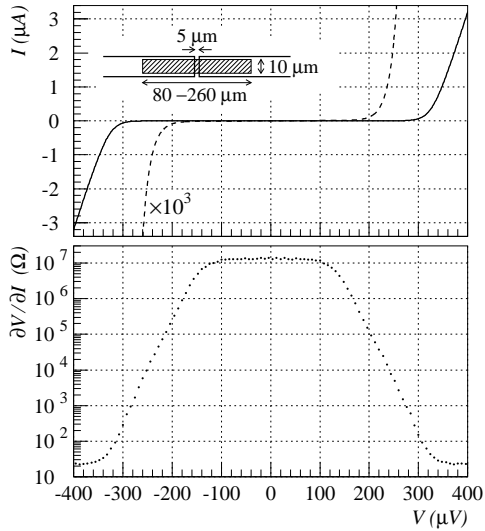}
   \caption{
   Measured current-voltage characteristics
   (solid line top graph) and corresponding
   differential resistance (bottom) of a high $\rho_n$ SINIS device,
   a schematic drawing of which is shown in the inset.
   The dashed line (top graph) represents the same data but with
   current scaled by $10^3$. }
   \label{IV.fig}
\end{figure}

We have fabricated high quality SINIS junctions by standard optical
lithography and metal evaporation deposition
(see inset of Fig.~\ref{IV.fig} for schematic drawing).
The superconducting electrodes
were pure Aluminium (Al, $\sim \!300 \, \mathrm{nm}$ thick),
whereas the normal absorber was Al
doped with 0.3--0.6\% of Manganese
(Mn, 10--500$ \, \mathrm{nm}$ thick)
in order to suppress Cooper pairing \cite{Ruggiero2004}.
An additional Silver layer
(0.1--8$ \, \mu \mathrm{m}$) was
eventually deposited on top of the AlMn absorber in order to increase
absorption efficiency to X-rays.
The native AlO$_\mathrm{x}$ forming the tunnel barrier yielded excellent
and highly stable device
characteristics with very low leakage currents.
The product of (single) junction area $A$ and normal resistance $R_n$ was
$\rho_n = R_n A = 0.6$--$40 \, \mathrm{k} \Omega \mu \mathrm{m}^2$
for different oxide thicknesses.
Figure~\ref{IV.fig} shows current-voltage characteristics of a high
$\rho_n$ device together with its differential resistance
$R_d = \partial V / \partial I$. The measured gap corresponds to twice the
Al gap ($\Delta_\mathrm{Al} = 172 \, \mu \mathrm{eV}$). This fact and
the perfect symmetry of the characteristics indicate no significant influence
of the Mn doping on tunneling current.
Low leakage is required for current biased detector operation.
From extra\-polation of the voltage dependent low current
$R_d$ to $V \rightarrow 0$
we extract a device base temperature of $T_b = 88 \, \mathrm{mK}$,
in spite of a cryostat temperature of about $50 \, \mathrm{mK}$.
The elevated $T_b$ is due to background blackbody radiation and
the presence of a relatively powerful $^{55}$Fe X-ray source.
The ratio of leakage resistance to $R_n$ was $5.6 \cdot 10^5$ for the
device in Fig.~\ref{IV.fig} (representative for high $\rho_n$ samples)
and decreased moderately for increasing barrier transparency
(lower $\rho_n$).

\begin{figure}[h]
   \centering
   \includegraphics[width=0.96\linewidth]{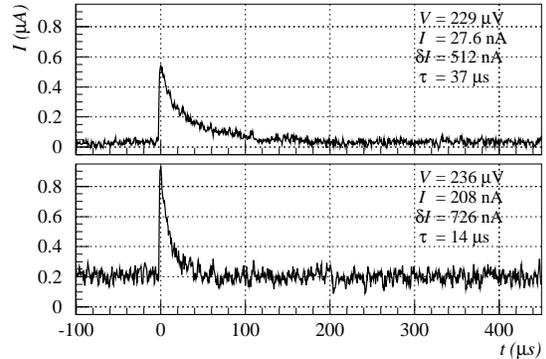}
   \caption{
   Measured SINIS detector signals from X-ray events at two
   different operation points.
   Note the change in signal dynamics (decay time $\tau$). }
   \label{signals.fig}
\end{figure}

Upon energy deposition in the absorber and (presumably quick)
energy transfer to the electron system the heat flow is dominated by
\cite{Jochum1998}:
\begin{itemize}
\item Electron-phonon coupling. The (hot) electrons relax to the (cold)
phonon bath with a rate \cite{Wellstood1994}
\begin{equation}
P_\mathrm{e-ph} = \Sigma \nu (T_e^5 - T_b^5) \, ,
\label{P-e-ph.eq}
\end{equation}
where
$\Sigma \approx 3 \, \mathrm{nW} \, \mathrm{K}^{-5}  \mu \mathrm{m}^{-3}$
is a material dependent coupling constant and $\nu$ is the absorber volume.
\item Hot-electron tunneling. The excess quasiparticles excited
above the gap are extracted yielding the desired current signal.
Power is removed irrespective of electrical current flow direction
(i.e.\ through both junctions) and is approximately given (per junction) by
\begin{equation}
P_\mathrm{tun} \approx \frac{I}{e} \, \max (\Delta -eV , kT) \, .
\label{P-tun.eq}
\end{equation}
\item Backheating. Excitations which are not efficiently removed from
the barrier region can leak back to the absorber volume \cite{Jochum1998}.
\item Power load from background radiation sources.
Here we estimate $P_\mathrm{bg} \approx 160 \, \mathrm{pW}$.
\end{itemize}
The thermal conductances $G = \d P / \d T$ determine the time constants
$\tau = C(T)/G$ of the relaxation processes, where $C(T) = \gamma T \nu$
is the electronic specific heat and $\gamma$ the Sommerfeld constant.
We have investigated the dynamics of detector signals due to X-ray events
as shown in Fig.~\ref{signals.fig} for two different bias voltages.

\begin{figure}[ht]
   \centering
   \includegraphics[width=0.85\linewidth]{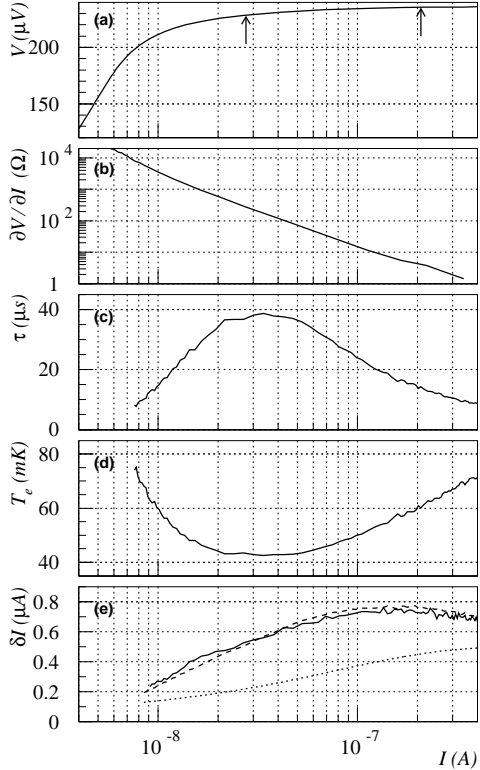}
   \caption{
   (a) Measured $IV$ characteristics of a low $\rho_n$
   SINIS device at $88 \, \mathrm{mK}$.
   The deviation from exponential behaviour at low currents
   ($<10 \, \mathrm{nA}$) is due to leakage onset.
   The arrows indicate the operation points for the signals
   shown in Fig.~\ref{signals.fig}.
   The abscissa was converted to display current
   (instead of bias voltage)
   for improved visibility in the relevant range.
   (b) Corresponding differential resistance.
   (c) Decay time $\tau$ of measured detector signals upon
   deposition of $6 \, \mathrm{keV}$  X-rays.
   (d) Deduced effective electron base temperature $T_e$
   of the absorber.
   (e) Measured current signal amplitude (solid line).
   The dashed line is a theoretical calculation assuming
   an effective $T_e$ as shown in (d), whereas the dotted
   line assumes $T_e = 88 \, \mathrm{mK}$. }
   \label{coolresults.fig}
\end{figure}

For this type of experiment we have typically chosen low $\rho_n$
devices since the self-cooling power is predicted \cite{Nahum1994}
to increase with increasing barrier transparency,
i.e.\  $P_\mathrm{tun} \propto R_n^{-1}$.
However, the junctions with the thinnest oxide barriers revealed a
partial suppression of the gap on the superconducting electrode
side due to  proximity of the ferromagnetic impurities
(see $IV$ in Fig.~\ref{coolresults.fig}a).
In order to keep the X-ray induced $T$ and
$I$ variations small to maintain the simple model approach,
a rather large ($5 \, \mu \mathrm{m}$ thick)
absorber volume was used.
Detector signals were recorded over the
bias voltage range of sufficient sensitivity
and fitted to the analytical model to extract
the (essentially exponential) pulse decay time $\tau$,
which is plotted in Fig.~\ref{coolresults.fig}c.

Using Eqs.~(\ref{P-e-ph.eq},\ref{P-tun.eq}) and neglecting
the effect of backheating we can calculate an effective
$T_e$ from $\tau$ and for given device parameters,
as shown in Fig.~\ref{coolresults.fig}d.
Due to its $T_e^5$ dependence the value of $T_e$ is not very
sensitive to modest variations of model parameters.
In Fig.~\ref{coolresults.fig}e we plot the measured
X-ray event signal amplitudes (solid line) together with
theoretical calculations using an electron tem\-pera\-ture
as determined in Fig.~\ref{coolresults.fig}d (dashed line)
as well as for $T_e = 88 \, \mathrm{mK}$ (dotted line).
Note the excellent agreement between experimental data
and theory assuming a variable $T_e$, and the increase
in sensitivity compared to a model without self-cooling.
The effect of microrefrigeration
compensates (in our case) at least for
the power load from background radiation.

In spite of the remarkable consistency between
measurements and our simple model, degradation due to
backheating should in most cases be con\-sidered.
Technically, the effect can be reduced by very thick
electrodes or implementation of trapping layers
\cite{Pekola2000}.
Furthermore, we observed an indication of incomplete
thermalization and partial phonon escape from the absorber.
This was reflected by a relatively poor spectral energy
resolution of our devices.
Improvements are expected from deposition
of the detector on a membrane \cite{Nahum1995}
or fabrication of a fully suspended absorber bridge
with small junction areas.

\section*{Acknowledgements}
We are grateful to Ph.\ Lerch for valuable discussions
and to F.\ Burri for technical support.


\begin{thebibliography}{00}
\bibitem{Nahum1995} M.\ Nahum and J.M.\ Martinis,
	Appl.\ Phys.\ Lett.\  \textbf{66} (1995) 3203.
\bibitem{Edwards1993} H.L.\ Edwards, Q.\ Niu, and A.L.\ de~Lozanne,
	Appl.\ Phys.\ Lett.\  \textbf{63} (1993) 1815.
\bibitem{Nahum1994} M.\ Nahum, T.M.\ Eiles, and J.M.\ Martinis,
	Appl.\ Phys.\ Lett.\  \textbf{65} (1994) 3123.
\bibitem{Golubev2001} D.\ Golubev and L.\ Kuzmin,
	J.\ Appl.\ Phys.\  \textbf{89} (2001) 6464.
\bibitem{Leivo1996} M.M.\ Leivo, J.P.\ Pekola, and D.V.\ Averin,
	Appl.\ Phys.\ Lett.\  \textbf{68} (1996) 1996.
\bibitem{Ruggiero2004} S.T.\ Ruggiero et al.,
	J.\ Low Temp.\ Phys.\  \textbf{134} (2004) 973.
\bibitem{Jochum1998}  J.\ Jochum et al.,
	J.\ Appl.\ Phys.\  \textbf{83} (1998) 3217.
\bibitem{Wellstood1994} F.C.\ Wellstood, C.\ Urbina, and J.\ Clarke,
	Phys.\ Rev.\ B  \textbf{49} (1994) 5942.
\bibitem{Pekola2000} J.P.\ Pekola et al.,
	Appl.\ Phys.\ Lett.\  \textbf{76} (2000) 2782.
\end{thebibliography}
\end{document}